# Public discourse and sentiment during the COVID 19 pandemic: using Latent Dirichlet Allocation for topic modeling on Twitter

Jia Xue[1,2], Junxiang Chen[3], Chen Chen[4], Chengda Zheng[2], Sijia Li[5, 6], Tingshao Zhu[5*]

[1] Factor-Inwentash Faculty of Social Work, University of Toronto, Toronto, Canada

[2] Faculty of Information, University of Toronto, Toronto, Canada

[3] School of Medicine, University of Pittsburgh, Pittsburgh, Pennsylvania, United States

[4] Middleware system research group, University of Toronto, Toronto, Canada

[5] Institute of Psychology, Chinese Academy of Sciences, Beijing, China

[6] Department of Psychology, University of Chinese Academy of Sciences, Beijing, China

*Corresponding author

E-mail: tszhu@psych.ac.cn (TSZ)

# Abstract

The study aims to understand Twitter users' discourse and psychological reactions to COVID-19. We use machine learning techniques to analyze about 1.9 million Tweets (written in English) related to coronavirus collected from January 23 to March 7, 2020. A total of salient 11 topics are identified and then categorized into ten themes, including "updates about confirmed cases," "COVID-19 related death," "cases outside China (worldwide)," "COVID-19 outbreak in South Korea," "early signs of the outbreak in New York," "Diamond Princess cruise," "economic impact," "Preventive measures," "authorities," and "supply chain." Results do not reveal treatments and symptoms related messages as prevalent topics on Twitter. Sentiment analysis shows that fear for the unknown nature of the coronavirus is dominant in all topics. Implications and limitations of the study are also discussed.

# Introduction

WHO declares COVID-19 as a global health pandemic. Social media has played a crucial role before the virus outbreak and continues to do so as it spreads globally. After China took strict quarantine measures as an intervention (e.g., cities on locked down, school closure, and employed self-isolation), Chinese social media platforms (e.g., Weibo, WeChat, Toutiao) become the lifeline for almost all isolated people who have been housebound for 30+ days and relying on these channels to obtain information, exchange opinions, socialize, and order food [1]. Existing studies [2-5] show that Twitter data can provide useful information for epidemic disease (e.g., H1N1, Ebola), including tracking rapidly evolving public sentiments, measuring public interests and concerns, estimating real-time disease activity and trends, and tracking reported disease levels.

However, these studies have limitations, with only qualitatively manual coding a very small number of Tweets. They require more advanced techniques to improve accuracy and precision for examining public opinions and sentiments. In addition, it remains unknown about public reactions to the COVID online. The vast majority of searched articles about COVID-19 and 2019-nCoV focus on epidemic control, such as the transmissibility of the virus [6], clinical characteristics of the infected cases [7], and patient screening [8].

The present study uses tremendous amounts of collected Twitter data to respond and add knowledge to our understandings of the pandemic. Aiming to explore the public discourse and psychological reactions during the early stage of COVID-19, we use a machine learning approach to examine (1) What latent topics related to COVID-19 can we identify from these Tweets? (2) What are the themes of these identified topics? (3) How Twitter users emotionally react to COVID-19 pandemic? And (4) How do these sentiments change over time?

# Methods

## Research design

We used an observational study design and a purposive sampling approach to select all the Tweets contained defined hashtags (e.g., #2019nCoV) related to COVID-19 on Twitter. We used natural language processing methods to find salient topics and terms related to COVID-19. Our Twitter data mining approach included data preparation and data analysis. Data preparation consisted of three steps: (1) sampling; (2) data collection; and (3) pre-processing the raw data. After pre-processing the raw dataset, we proceeded to the data analysis stage, including (1) unsupervised machine learning, (2) qualitative method; and (3) sentiment analysis. The unit of analysis was each message-level Tweet posted on Twitter.

## Sampling and data collection

We purposely selected a list of 19 trending hashtags related to COVID-19 as key search terms to collect Tweets on Twitter (S1 Tables). We used Twitter's open application programming interface (API) to collect Tweets published between January 23, 2020, and March 7, 2020. We used the Python code provided by Twitter Developer [9] to access the Twitter API. Shown in Fig 1, a total of 20 million (n=20,370,854) Tweets were collected. After we removed the non-English Tweets (n=9,694,320), duplicates and retweets (n=7,731,035), 1.9 million (n=1,963,285) Tweets were our dataset for this study. The following features were collected for each single Tweet message (1) each message-level tweets (full text); (2) function features of (a) hashtags; (b) the number of favorites; (c) the number of followers; (d) the number of friends; (e) number of retweets; (f) user location; and (g) user description.

**Fig 1. Preprocessing data chart**

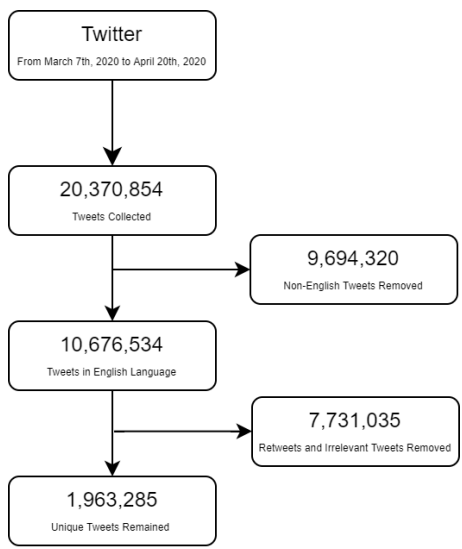

## Pre-processing the raw dataset

We pre-processed the raw data to ensure quality. We used Python, a programming language, to conduct data analysis. The pre-processing plan was as follows:

(1) We removed the hashtag symbol and its content (e.g., #COVID19), @users, and URLs from the messages because the hashtag symbols or the URLs did not contribute to the message analysis.

(2) We removed all non-English characters (non-ASCII characters) because the study focused on the analysis of messages in English.

(3) We removed repeated words. For example, sooooo terrified was converted to so terrified.

(4) We removed special characters, punctuations, and numbers from the dataset as they did not help with detecting the profanity comments.

## Data analysis

### Unsupervised machine learning

We used unsupervised machine learning to examine data for patterns because this approach was commonly used when studies had little observations or insights of the unstructured text data. A qualitative approach had challenges analyzing the large scale of Twitter data. Unsupervised learning derived a probabilistic clustering based on the data itself, allowing us to conduct exploratory analyses of large unstructured texts in social science research. We configured topic modeling, an unsupervised machine learning method, to generate top latent topic distributions. Latent Dirichlet allocation (LDA) [10] was a probabilistic model of word counts that analyzes a set of documents. We used LDA to identify patterns, themes, and structures of the Tweets texts and examine how these themes were connected. It enabled us to efficiently categorize the large bodies of data based on patterns and features. LDA had been used to do sentiment analysis of Tweets related to health [11]. Topic modeling had been widely used to gain a descriptive understating of unstructured Twitter big data in social science research [12].

**Qualitative analysis**

We triangulated and contextualized findings from unsupervised learning in the study. We employed the qualitative approach to support deeper qualitative dives into the dataset, such as labeling popular words and Tweet topics, assigning meanings and themes to the topics, interpreting the themes and patterns identified from the Tweets [13], and inductively developing themes for the latent topics generated by machine algorithms. The qualitative approach relies on the diverse, in-depth interpretations from human, which allows for inductive, exploratory analysis, and the application of theoretical approaches [14].

**Sentiment analysis**

Sentiment analysis was a computational and natural language processing-based method that analyzed the people's sentiment, emotions, and attitudes in given texts [15] and an essential method in social media research. The sentiment analysis in the present study was based on a machine learning model for predicting emotions from English Tweets [16]. This model classified each tweet into eight pairwise emotions in Plutchik's wheel of emotions [17], including joy-sadness, trust-disgust, fear-anger, and surprise-anticipation. This method returned one emotion from the eight categories for each given Tweet.

# Results

## Descriptive results

After pre-processing the collected tweets, our final dataset consisted of 1,963,285 Tweets after removing the duplicates mentioning at least one of the nineteen hashtags from January 23 to March 7, 2020. Fig 2 presented the number of Tweets under the top 9 hashtag by dates ("#Coronavirus", n = 1,405,254, "#Wuhan", n = 144,240, "#Wuhancoronavirus", n = 73,393, "#Coronaoutbreak", n = 73,147, "#2019ncov", n = 60,278, "#ChinaCoronavirus", n = 19,188, "#Chinavirus", n = 17,865, "#CoronavirusChina", n = 16,371, "#Wuhanoutbreak", n = 10,548). The number of Tweets using hashtag #coronavirus gradually increased since February 14 and dropped on March 1 when the hashtag #wuhanoutbreak suddenly increased for four days.

**Fig 2. The number of Tweets under the top 9 hashtags by dates.**

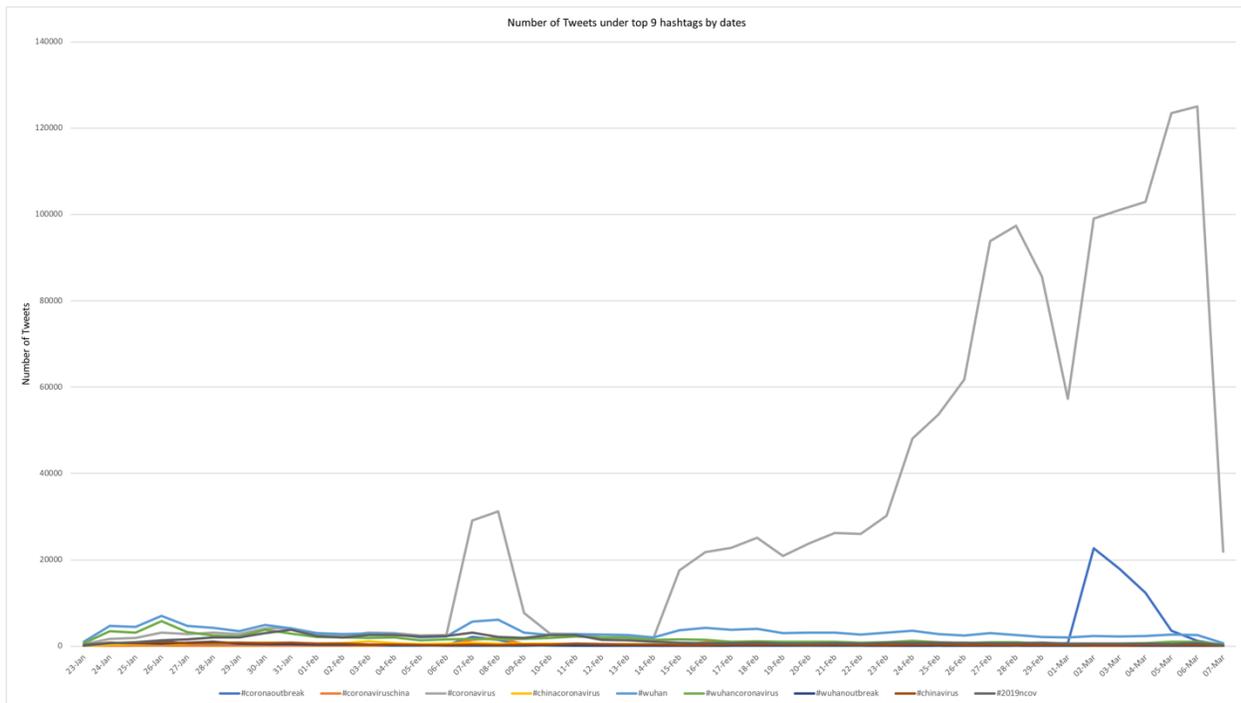

## COVID-19 related topics

The automated machine learning LDA approach generated commonly co-occurred words and also organized them into different topics. We calculated the most appropriate number of topics based on the coherence model – gensim [18]. We chose the number of topics to be 11 returned by LDA for this dataset because it had the highest coherence score. Fig 3 showed the coherence score for the number of topics returned by the LDA model.

**Fig 3. Coherence score for the number of topics**

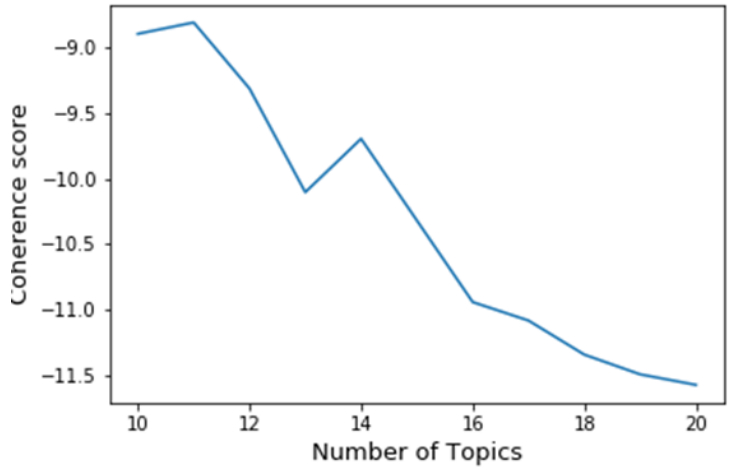

We analyzed the document-term matrix with the chosen 11 topics and obtained the distributions of the 11 topics. Table 1 presented the results of identified 11 salient topics, the most popular pairs of words within each topic, and the number of Tweets under each topic.

**Table 1. Identified salient topics and their components (bi-grams).**

| Topic | Bigrams within topics | Numbers of Tweets |
|---|---|---|
| 1 | toilet paper, w/ecosearch, ecosearch news, news web, health emergency, corona virus, fake news, xi jinping, dont want, self isolate, want know, covid 19, number people, breaking news, read here, good idea, health officials, spanish flu, new York | 334,193 |
| 2 | diamond princess, disease control, donald trump, li wenliang, covid 19, tests positive, dr li, corona virus, common cold, shaking hands, south korea, details gt, hong kong, supply chains, tested positive, centers disease, control prevention, president trump, supply chain | 158,704 |
| 3 | face masks, social media, people die, new York, panic buying, corona virus, 1 1, loved ones, coronavirus outbreak, case confirmed, watch video, u s, tested positive, million people, medical staff, like this, shake hands, high school, coronavirus update | 161,361 |
| 4 | u s, death rate, mortality rate, mike pence, 3 4, coronavirus death, toll rises, fatality rate, white house, dont know, confirms case, south korea, rate 3, chief medical, coronavirus spread, public health, medical officer, china coronavirus, climate change | 160,237 |
| 5 | tested positive, outside china, total cases, sars cov, cov 2, cases confirmed, year old, date total, coronavirus case, north korea, deaths date, confirmed worldwide, covid 19, infectious disease, new case, positive case, communist party, confirmed cases | 145,781 |
| 6 | coronavirus outbreak, year old, gt gt, covid 19, thank you, amid coronavirus, confirmed case, wall street, economic impact, united states, travel ban, good news, stock market, amp a, press conference, q amp, whats happening, corona virus, years old | 152,724 |
| 7 | washing hands, south korea, prevent spread, 2019 ncov, test kits, covid 19, novel coronavirus, 20 seconds, need know, stock market, soap water, happy birthday, 2019 novel, 1 000, coronavirus outbreak, coronavirus cases, help prevent, hands soap, reported today | 151,935 |
| 8 | diamond princess, stay home, 14 days, work home, princess cruise, tested positive, looks like, face mask, test positive, task force, supply chain, san Francisco, wearing masks, corona virus, coronavirus fears, hong kong, ship japan, dont forget, 14 day | 165,730 |
| 9 | world health, health minister, health organization, press conference, washington state, coronavirus covid, live updates, tested positive, number cases, state emergency, 19 cases, new York, 19 outbreak, bbc news, health ministry, people died, right now, coronavirus disease, novel coronavirus | 162,623 |
| 10 | stay safe, corona virus, stop spread, chinese people, corona beer, infectious diseases, s amp, ive seen, dont know, health minister, health crisis, worst case, good thing, god bless, amp p, case scenario, pence charge, help stop, im worried | 157,064 |
| 11 | confirmed cases, south korea, bringing total, cases confirmed, total confirmed, cases reported, total number, new deaths, total deaths, wash hands, coronavirus cases, number confirmed, touch face, cases coronavirus, hubei province, number cases, 2 new, cases bringing, new confirmed | 170,834 |

## COVID-19 related themes

We generated some representative Tweets on each topic to explain the themes of these topics. Two authors discussed the bigrams and representative Tweets in each of the 11 topics and then categorized them into ten themes (Table 2). In addition, we computed the topic distance [10] and presented a 2D plane of the intertopic distance [19] in Fig 4. Each circle represented a topic from Topic 1 to Topic 13 in the study. The centers are determined by computing the distance between topics. In the visualization, these circles were not overlapped, which cross-validated the classification of the ten themes.

**Fig 4 Intertopic distance map**

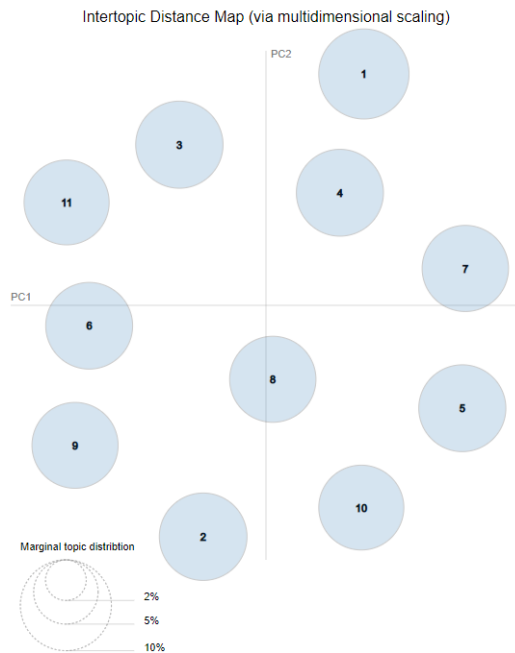

*Notes: PS: principal component. The circled areas are the overall prevalence, and the circle center is determined by computing the distance between topics.*

Table 2 presented the identified topics and themes, and each row of bigrams represented one topic under the theme. We identified ten themes, such as "updates about the number of COVID-19 cases (*confirmed cases*, *total confirmed*, *cases reported*)," "COVID-19 related death [(*new deaths*, total deaths) and (*people die*, death rates)]," and "preventive measures [(*toilet paper*, *self-isolate*), (*face masks*, *panic buying*), travel bans, and (*washing hands*, *test kits*, *20 seconds*, *soap water*, *hands soap*)]".

**Table 2. Themes and most likely topic components (bi-grams).**

| Theme | Bigrams within topics |
|---|---|
| Updates about the number of COVID-19 cases | • confirmed cases, total confirmed, cases reported, total number, number confirmed, new confirmed<br>• tested positive, cases confirmed, new case, positive case |
| COVID-19 related death | • new deaths, total deaths<br>• death rate, mortality rate, fatality rate, coronavirus spread<br>• people die |
| Cases outside China | • outside china, confirmed worldwide<br>• Hong Kong<br>• ship Japan |
| Outbreak in South Korea | • South Korea, 2019 ncov, covid19, novel coronavirus |
| Early signs of the outbreak in New York city | • health emergency, corona virus, fake news, want know, covid 19, New York<br>• people die, New York, panic buying, corona virus, case confirmed, tested positive, high school |
| Diamond princess cruise | • diamond princess, disease control, tests positive<br>• princess cruise, ship japan |
| Economic impact | • wall street, economic impact, united states, stock market |
| Preventive measures | • toilet paper, self-isolate<br>• shaking hands, control prevention<br>• face masks<br>• travel ban<br>• washing hands, test kits, 20 seconds, soap water, hands soap<br>• stay home, 14 days, work home, wearing masks |
| Authorities | • Xi jinping, health officials<br>• disease control, Donald Trump, President Trump<br>• Li wenliang, dr li<br>• medical staff<br>• white house, chief medical, public health, medical officer<br>• North Korea, communist party<br>• health minister, health organization, Washington state<br>• Mike pence |
| Supply chain | • supply chains<br>• panic buying |

Table 3 highlighted the representative Tweets within each topic under each theme. To protect the privacy and anonymity of the Twitter users of these sample Tweets, we used either excerpt of Tweets or paraphrased several terms in the message.

**Table 3. Representative Tweets within themes.**

| Theme | Tweets samples |
|---|---|
| Updates about the number of COVID-19 cases | • "…over 5,000 cases of confirmed #COVID19 …"<br>• "…there are 101,765 confirmed cases of the coronavirus …"<br>• "…47,885 recovered (+2,270)…" |
| COVID-19 related death | • "@healthdirectAU: there are currently 33 confirmed cases of coronavirus in Australia…"<br>• "… coronavirus… and 3,461 deaths globally…"<br>• "…US has near 10% death rate from #coronavirus…" |
| Cases outside China (worldwide) | • "…beyond China, total confirmed cases reach 4,154 as of Feb.27th …"<br>• "…#covid19 is now in 50 countries/regions… several countries declared their confirmed cases of covid…"<br>• "…excluding #China: 10,283 confirmed, 792 recovered, 173 deaths…" |
| Outbreak in South Korea | • "…a vast majority of coronavirus patients in Korea are linked to the Shincheonji church…"<br>• "...South Korean city face shortage of hospital bed as #outbreak expands…"<br>• "#southkorea declares 'war' on #coronavirus …" |
| Early signs of the outbreak in New York city | • "…in the news, NYC orders mandatory coronavirus testing for public workers …"<br>• "@homedepot,@lowes, and any respectable hardware store from the bottom of NYC all the way upstate to Rochster is completely sold out of all respiratory masks…" |
| Diamond princess cruise | • "…approx..100 more people on Princess Diamond showed symptoms like a fever, and will be tested soon…"<br>• "…passenger of Diamond Princess ship tested positive for the virus #2019nCoV…"<br>• "…61 people now infected on #DiamondPrincess cruise ship off japan #coronavirus…" |
| Economic impact | • "…IMF chief says the outbreak could derail global economic growth…"<br>• "... https://t.co/OtsbHOZBTW #economicoutlook #markets<br>• #globaleconomy #Coronavirus likely to impact…"<br>• "…airline stocks crash, face turbulence amid coronavirus…airline stocks fell significantly on Thursday …" |
| Preventive measures | • "…a crappy coronavirus shortage toilet paper …"<br>• "…my understanding is that the best way to stop the spread of #covid19 is to use hand sanitizer and not touch my face…" |
| | - "…stay safe wearing masks, avoid outside plans, stay at home as much as you can #coronavirusoutbreak…"<br>- "…we've had travel bans for over 4 weeks…" |
| Authorities | - "… Trump lied about #coronavirus, vote him out #voteblue #JoeBiden2020…"<br>- "coronavirus 'likely' to hit UK – professors say public health officials must do more #coronavirus…"<br>- "Mike pence will stop #coronavirus with gender segregated workplaces and don't tell him otherwise…"<br>- "…Chinese doctor #LiWenLiang, one of the eight HERO whistleblowers who tried to warn other …"<br>- "…is the the figure #WHO told us the coronavirus is under control? Let there be no panic…"<br>- "…the PRESIDENT OF THE UNITED STATES said the coronavirus was not a concern anymore #CDC…" |
| Supply chain | - "with #wuhancoronavirus, the supply chain in China will soon collapse, better prepare for the global shortage of supply of everything…?<br>- "…@Catalysis3D can help with low cost and fast additive manufactured bridge tooling and part…#supplychain…"<br>- "…companies re-evaluating supply chains due to #coronavirus… let's revisit how #PLM can help…" |

## Sentiment analysis

Tweets contained information about people's thoughts and emotions [20]. We presented individuals' emotional reactions to the COVID-19 pandemic in Fig 5. It represented the proportion of emotional tweets over daily tweets by date. *Fear* (yellow line) was consistently the dominant emotion over time, which was about 50% of daily Tweets from the Wuhan outbreak to early March. Proportionally lower than feeling of fear, Tweets on trust (brown line) slightly increased over time.

**Fig 5. Emotions trends during the early stages of the COVID-19**

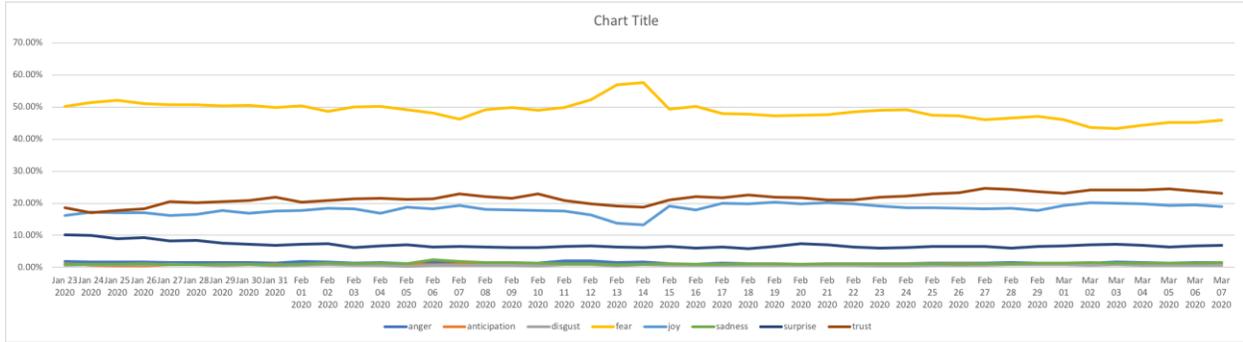

# Sentiments within 11 topics

Table 4 showed the percentage of each emotion within each of the 11 topics. Across all topics, we observed that the feeling of fear has been prominent. For example, fear for the unknown nature of the COVID-19 consisted of almost 50% of the Tweets in all eleven topics. Approximately 24% of the emotions within Tweets under Topic 1 related to the public's *trust* for the health authorities.

Since fear was prominent in all eleven topics, we further ran a one-tailed *z* test and assessed if each of the eight emotions was statistically significantly different across topics. We used a *p-value* smaller than .001 as a threshold and presented the results in Table 4. For example, fear for the uncertainty about COVID-19 was found to have a higher probability of being prevalent in Topics 1, 4, 9, and 11. Trust expressed in Tweets was statistically significant prevalent in Topics 1, 2, and 10. Surprise for the pandemic was statistically significant frequent in Topics 1 and 11. Joy was statistically significant widespread in Topics of 5, 7, 8, and 11.

**Table 4. Percentage of each emotion within 11 topics and *p-value* from Z-test**

|         | Anger | Anticipation | Disgust | Fear      | Joy   | Sadness | Surprise | Trust      |
|---------|-------|--------------|---------|-----------|-------|---------|----------|------------|
| Topic 1 | 1.3%  | 1.2%         | 0.6%    | 47.6%***  | 15.5% | 1.5%    | 8.7%***  | 23.7%***   |
| Topic 2 | 1.4%  | 1.1%         | 0.8%    | 45.3%     | 19.5% | 1.4%    | 6.3%     | 24.2%***   |
| Topic 3 | 1.5% | 1.2% | 0.8% | 46.3% | 19.6% | 1.4% | 6.1% | 23.1% |
| Topic 4 | 1.5% | 1.1% | 0.8% | 47.4%*** | 18.9% | 1.3% | 6.1% | 23.1% |
| Topic 5 | 1.3% | 1.2% | 0.8% | 45.9% | 20.9%*** | 1.3% | 6.2% | 22.4% |
| Topic 6 | 1.6% | 1.1% | 0.8% | 45.8% | 19.7% | 1.2% | 6.0% | 23.8% |
| Topic 7 | 1.6% | 1.0% | 0.7% | 45.8% | 20.3%*** | 1.2% | 5.8% | 23.6% |
| Topic 8 | 1.6% | 1.1% | 0.8% | 44.9% | 20.5%*** | 1.2% | 6.3% | 23.5% |
| Topic 9 | 1.3% | 1.1% | 0.7% | 47.6%*** | 19.3% | 1.3% | 6.2% | 22.6% |
| Topic 10 | 1.6% | 1.1% | 0.9% | 44.9% | 19.4% | 1.4% | 5.9% | 24.7%*** |
| Topic 11 | 1.1% | 1.4% | 0.6% | 47.7%*** | 20.4%*** | 1.1% | 7.2%*** | 20.6% |

Notes: The sum of the percentage under each topic is not equal to 100%. The rests are either neutral or other emotions. *** $p < .001$

## Discussion and conclusion

This study shows Twitter users' discussions and sentiments to the COVID-19 from January 23 to March 7, 2020. Our findings facilitate an understanding of public discussions and sentiments to the outbreak of COVID-19 in a rapid and real-time way, contributing to the surveillance system to understand the evolving situation. The study overcomes the limitations of the traditional social science approach, which relies on time-consuming, retrospective, time-lagged, small-scale surveys, and interviews. The identified patterns and emotions of public tweets could be used to guide targeted intervention programs.

First, early recognition of COVID-19 cases and a potential outbreak in New York City were identified among a massive number of tweets, suggesting that the Twitter community has acknowledged the disease severity as early as February. A small peak of the Tweets volume is identified between Feb 10th and 14th, and then gradually increase again after Feb.14th. This finding is also timed with the very first CDC's warning on Twitter (@CDCgov) on February 10, 2020: "If you've recently from China, know the symptoms of #2019nCoV. These include mild to severe respiratory illness with fever, cough, shortness of breath. See bit.ly/38zjnYo." An increasing number of Tweets may be followed with CDC's post, suggesting a good opportunity to guide the public to take action to take preventive measures in February. Rapidly identifying and utilizing social media messages may help the public and authorities to respond to the spread of the disease at the early stages.

Second, discussions of COVID-19 symptoms (e.g., cough, fever, difficulty breathing) and treatments (e.g., vaccine, rest and sleep, drink liquids) were notably missing from our collected Tweets from January 23 to March 7, 2020. One study selects Tweets (n=35,786) associated with COVID-19 symptoms (e.g., diagnosed, pneumonia, fever, cough) from March 3 to 20, 2020, and finds that the volume of signal Tweets for symptoms increases over time [21]. The inconsistent findings suggest that Twitter is not widely used as a platform for posting symptoms or seeking medical help. Findings inform that more treatment-related messages can be posted as an educational tool for the public on social media Health authorities or public health communities.

Third, fear is a dominant emotion in all topics during the early stages of the COVID-19 pandemic. Results are consistent with other studies [22-25], which shows that COVID-19 significantly impacts individuals' psychological conditions. Sentiment analysis of the COVID-19 pandemic related content contributes to our understanding of the dynamics of online users' concerns and feelings during the epidemic. Our findings have implications for health authorities that mental health and psychosocial well-being support is needed during this time [20].

There are several limitations to the study. First, we only sample a trending of 19 hashtags as search terms to collect Twitter data. Some new hashtags have become new trending terms for Twitter users to group topics over time. For example, #COVID19 has been widely used after it becomes the official name for the virus. Second, Twitter users are not representative of the whole population and only indicate online users' opinions and reactions about COVID-19. However, the Twitter dataset is a valuable source for understanding the real-time Twitter user-generated content related to COVID-19 disease activities. Third, non-English Tweets are removed from the analysis, and results are limited to a particular population. Future studies are recommended to include Italian, Germany, and Spanish languages for COVID-19 analysis.

## Acknowledgments

Artificial Intelligence Lab for Justice at the University of Toronto supports the data collection.

# Supporting information

S1 Table. (Table Hashtags used as key search terms)

**S1 Table Hashtags used as key search terms**

| Hashtags used as key search terms | #Coronaoutbreak, #CoronavirusChina, #Wuhan, #Coronavirus, #ChinaCoronavirus, #Wuhan #WuhanCoronavirus, #Wuhanoutbreak, #ChinaVirus, #2019nCoV, #ChineseDon'tComeToJapan, #NoSoyUnVirus, #IamNotVirus, #JeNeSuisPasUnVirus, #Xenophobia, #PrayForChina, #DrLiWenLiang, #ItWillGetBetter, #BeStrongChina. |
|---|---|